\documentclass[english]{article}
\pdfoutput=1
\usepackage[T1]{fontenc}
\usepackage[latin9]{inputenc}
\usepackage{refstyle}
\usepackage{url}
\usepackage{enumitem}
\usepackage{amsmath}
\usepackage{graphicx}
\usepackage[authoryear]{natbib}

\makeatletter

\AtBeginDocument{\providecommand\eqref[1]{\ref{eq:#1}}}
\providecommand{\tabularnewline}{\\}
\RS@ifundefined{subsecref}
  {\newref{subsec}{name = \RSsectxt}}
  {}
\RS@ifundefined{thmref}
  {\def\RSthmtxt{theorem~}\newref{thm}{name = \RSthmtxt}}
  {}
\RS@ifundefined{lemref}
  {\def\RSlemtxt{lemma~}\newref{lem}{name = \RSlemtxt}}
  {}

\usepackage{mathptmx}
\usepackage[symbol]{footmisc}

\makeatother

\usepackage{babel}
\begin{document}
\title{\textsc{A}ssigning credit to scientific datasets using article citation
networks}
\author{Tong Zeng$^{1,2,}$\thanks{Both authors contributed equally to this work}
, Longfeng Wu$^{\ast,}$\thanks{The author performed this work while visiting the Science of Science
and Computational Discovery Lab in the School of Information Studies
at Syracuse University} , Sarah Bratt$^{2}$\\
Daniel E. Acuna$^{2,}$\thanks{Corresponding author: \protect\url{deacuna@syr.edu}}}
\date{$^{1}$School of Information Management, Nanjing University, Nanjing
210023, China\\
$^{2}$School of Information Studies, Syracuse University, Syracuse,
NY 13244, USA}
\maketitle
\begin{abstract}
A citation is a well-established mechanism for connecting scientific
artifacts. Citation networks are used by citation analysis for a variety
of reasons, prominently to give credit to scientists\textquoteright{}
work. However, because of current citation practices, scientists tend
to cite only publications, leaving out other types of artifacts such
as datasets. Datasets then do not get appropriate credit even though
they are increasingly reused and experimented with. We develop a network
flow measure, called \textsc{DataRank}, aimed at solving this gap.
\textsc{DataRank} assigns a relative value to each node in the network
based on how citations flow through the graph, differentiating publication
and dataset flow rates. We evaluate the quality of \textsc{DataRank}
by estimating its accuracy at predicting the usage of real datasets:
web visits to GenBank and downloads of Figshare datasets. We show
that \textsc{DataRank} is better at predicting this usage compared
to alternatives while offering additional interpretable outcomes.
We discuss improvements to citation behavior and algorithms to properly
track and assign credit to datasets.
\end{abstract}

\section{Introduction}

A citation network is an important source of analysis in science.
Citations serve multiple purposes such as crediting an idea, signaling
knowledge of the literature, or critiquing others' work \citep{martyn1975citation}.
When citations are thought of as impact, they inform tenure, promotion,
and hiring decisions \citep{meho2007impact}. Furthermore, scientists
themselves make decisions based on citations, such as which papers
to read and which articles to cite. Citation practices and infrastructures
are well-developed for journal articles and conference proceedings.
However, there is much less development for dataset citation. This
gap affects the increasingly important role that datasets play in
scientific reproducibility \citep{task2013out,belter2014measuring,robinson2016analyzing,park2018informal},
where studies use them to confirm or extend the results of other research
\citep{sieber1995not,darby2012enabling}. One historical cause of
this gap is the difficulty in archiving datasets. While less problematic
today, the citation practices for datasets take time to develop. Better
algorithmic approaches to track dataset usage could improve this state.
In this work, we hypothesize that a network flow algorithm could track
usage more effectively if it propagates publication and dataset citations
differently. With the implementation of this algorithm, then, it will
be possible to correct differences in citation behavior between these
two types of artifacts, increasing the importance of datasets as first
class citizens of science.

Different researchers use citation networks to evaluate the importance
of authors \citep{ding2009pagerank,ding2011applying,west2013author},
papers \citep{chen2007finding,ma2008bringing}, journals \citep{bollen2006journal,bergstrom2007eigenfactor},
institutions \citep{fiala2013suborganizations} and even countries
\citep{fiala2012bibliometric}. The \textsc{PageRank} algorithm \citep{page1999pagerank}
has served as a base for much of these citation network-based evaluations.
For example, \citet{bollen2006journal} proposed a weighted \textsc{PageRank
}to assess the prestige of journals, while \citet{ding2009pagerank}
and \citet{ding2011applying} proposed a weighted \textsc{PageRank
}to measure the prestige of authors. \citet{fiala2012time} defined
a time-aware \textsc{PageRank }method for accurately ranking the most
prominent computer scientists. \citet{franceschet2017timerank} introduced
an approach called \textsc{TimeRank} for rating scholars at different
time points. \textsc{TimeRank} updates the rating of scholars based
on the relative rating of the citing and cited scholars at the time
of the citation. Citation networks are thus an important source of
information for ranking homogenous types of nodes.

Historically, ranking datasets using citation networks is significantly
more challenging. These challenges have technical and social issues
alike. First, datasets cost time and labor to prepare and to share,
resulting in some articles failing to provide datasets \citep{alsheikh2011public}.
Second, archiving and searching massive datasets is prohibitively
expensive and difficult. Third, scholars are not used to citing datasets.
Survey research shows that scholars value citing dataset \citep{kratz2015researcher}
yet they tend to cite the \emph{article} rather than the dataset or
they merely mention the dataset without explicit reference \citep{force2014encouraging}.
Therefore, these reasons have prevented the proper assignment of credit
to dataset usage.

Several initiatives attempt to improve citation practices for datasets.
In 2014, the Joint Declaration Of Data Citation Principles was officially
released. These principles, however, mainly focus on normalizing dataset
references rather than normalizing storage and some other technical
issues \citep{altman2015introduction,callaghan2014joint,mooney2012anatomy}.
For instance, some researchers have suggested assigning specific DOIs
to datasets to mitigate differences between datasets and articles
\citep{callaghan2012making}. Others have proposed to automatically
identify uncited or unreferenced datasets used in articles \citep{boland2012identifying,kafkas2013database,ghavimi2016identifying}.
All these solutions try to make citation dataset behavior more standard
or attempt to fix the citation network by estimating which data nodes
are missing. Therefore, these solutions necessarily modify the source
that algorithms use to estimate impact.

In this article, we develop a method for assigning credit to datasets
from citation networks of publications, assuming that dataset citations
have biases. Importantly, our method does not modify the source data
for the algorithms. The method does not rely on scientists explicitly
citing datasets but infers their usage. We adapt the network flow
algorithm of \citet{walker2007ranking} by including two types of
nodes: datasets and publications. Our method simulates a random walker
that takes into account the differences between obsolescence rates
of publications and datasets, and estimates the score of each dataset---the
\textsc{DataRank}. We use the metadata from the National Center for
Bioinformatics (NCBI) GenBank nucleic acid sequence and Figshare datasets
to validate our method. We estimate the relative rank of the datasets
with the \textsc{DataRank} algorithm and cross-validate it by predicting
the actual usage of them---number of visits to the NCBI dataset web
pages and downloads of Figshare datasets. We show that our method
is better at predicting both types of usages compared to citations
and has other qualitative advantages compared to alternatives. We
discuss interpretations of our results and implications for data citation
infrastructures and future work.

\section{Why measure dataset impact?}

Scientists may be incentivized to adopt better and broader data sharing
behaviors if they, their peers, and institutions are able to measure
the impact of datasets (e.g., see \citet{silvello2018theory} and
\citet{kidwell2016badges}). In this context, we review impact assessment
conceptual frameworks and studies of usage statistics and crediting
of scientific works more specifically. These areas of study aim to
develop methods for scientific indicators of the usage and impact
of scholarly outputs. Impact assessment research also derives empirical
insights from research products by assessing the dynamics and structures
of connections between the outputs. These connections can inform better
policy-making for research data management, cyberinfrastructure implementation,
and funding allocation.

Methods for measuring usage and impact include a variety of different
dimensions of impact, from social media to code use and institutional
metrics. Several of these approaches recognize the artificial distinction
between the scientific process and product \citep{priem2013scholarship}.
For example, altmetrics is one way to measure engagement with diverse
research products and to estimate the impact of non-traditional outputs
\citep{priem2014altmetrics}. Researchers predict that it will soon
become a part of the citation infrastructure to routinely track and
value \textquotedblleft citations to an online lab notebook, contributions
to a software library, bookmarks to datasets from content-sharing
sites such as Pinterest and Delicious\textquotedblright{} \citep[from ][]{priem2014altmetrics}.
In short, if science has made a difference, it will show up in a multiplicity
of places. As such, a correspondingly wider range of metrics are needed
to attribute credit to the many locations where research works reflect
their value. For example, datasets contribute to thousands of papers
in NCBI\textquoteright s Gene Expression Omnibus and these attributions
will continue to accumulate, just like paper accumulate citations,
for a number of years after the datasets are publicly released \citep{piwowar2011data,piwowar2013altmetrics}.
Efforts to track these other sources of impact include ImpactStory,
statistics from FigShare, and Altmetric.com \citep{robinson2017datacite}.

Credit attribution efforts include those by federal agencies to expand
the definition of scientific works that are not publications. For
example, in 2013 the National Science Foundation (NSF) recognized
the importance of measuring scientific artifacts other than publications
by asking researchers for \textquotedblleft products\textquotedblright{}
rather than just publications . This represents a significant change
in how scientists are evaluated \citep{piwowar2013altmetrics}. Datasets,
software, and other non-traditional scientific works are now considered
by the NSF as legitimate contributions to the publication record.
Furthermore, real-time science is presented in several online mediums;
algorithms filter, rank, and disseminate scholarship as it happens.
In sum, the traditional journal article is increasingly being complemented
by other scientific products \citep{priem2013scholarship}.

Yet despite the crucial role of data in scientific discovery and innovation,
datasets do not get enough credit \citep{silvello2018theory}. If
credit was properly accrued, researchers and funding agencies would
use this credit to track and justify work and funding to support datasets---consider
the recent Rich Context Competition which aimed at to filling this
gap by detecting dataset mentions in full-text papers \citep{zengnyu}.
Because these dataset mentions are not tracked by current citation
networks, this leads to biases in dataset citations \citep{robinson2016analyzing}.
The FAIR (findable, accessible, interoperable, reproducible) principles
of open research data are one major initiative that is spearheading
better practices with tracking digital assets such as datasets \citep{wilkinson2016fair}.
However, the initiative is theoretical, and lacks technical implementation
for data usage and impact assessment. There remains a need to establish
methods to better estimate dataset usage.

\section{Materials and methods}

\subsection{Datasets}

\subsubsection{OpenCitations Index (COCI)}

The OpenCitations index (COCI) is an index of Crossref's open DOI-to-DOI
citation data. We obtained a snapshot of COCI (November 2018 version),
which contains approximately 450 million DOI-to-DOI citations. Specifically,
COCI contains information including citing DOI, cited DOI, the publication
date of citing DOI. The algorithm we proposed in the paper requires
the publication year. However, not all the DOIs in COCI have a publication
date. We will introduce Microsoft Academic Graph to fill this gap.

\subsubsection{Microsoft Academic Graph (MAG)}

The Microsoft Academic Graph is a large heterogeneous graph consisting
of six types of entities: paper, author, institution, venue, event,
and field of study \citep{sinha2015overview}. Concretely, the description
of a paper consists of DOI, title, abstract, published year, among
other fields. We downloaded a copy of MAG in November 2019, which
contains 208,915,369 papers. As a supplement to COCI, we extract the
DOI and published year from MAG to extend those DOIs in COCI without
a publication date.

\subsubsection{PMC Open Access Subset (PMC-OAS)}

The PubMed Central Open Access Subset is a collection of full-text
open access journal articles in biomedical and life sciences. We obtained
a snapshot of PMC-OAS in August 2019. It consists of about 2.5 million
full-text articles organized in well-structured XML files. The articles
are identified by a unique id called PMID. We also obtained a mapping
between PMIDs and DOIs from NCBI, which enabled us to integrate PMC-OAS
into the citation network.

\subsubsection{GenBank}

\begin{figure}
\begin{centering}
\includegraphics{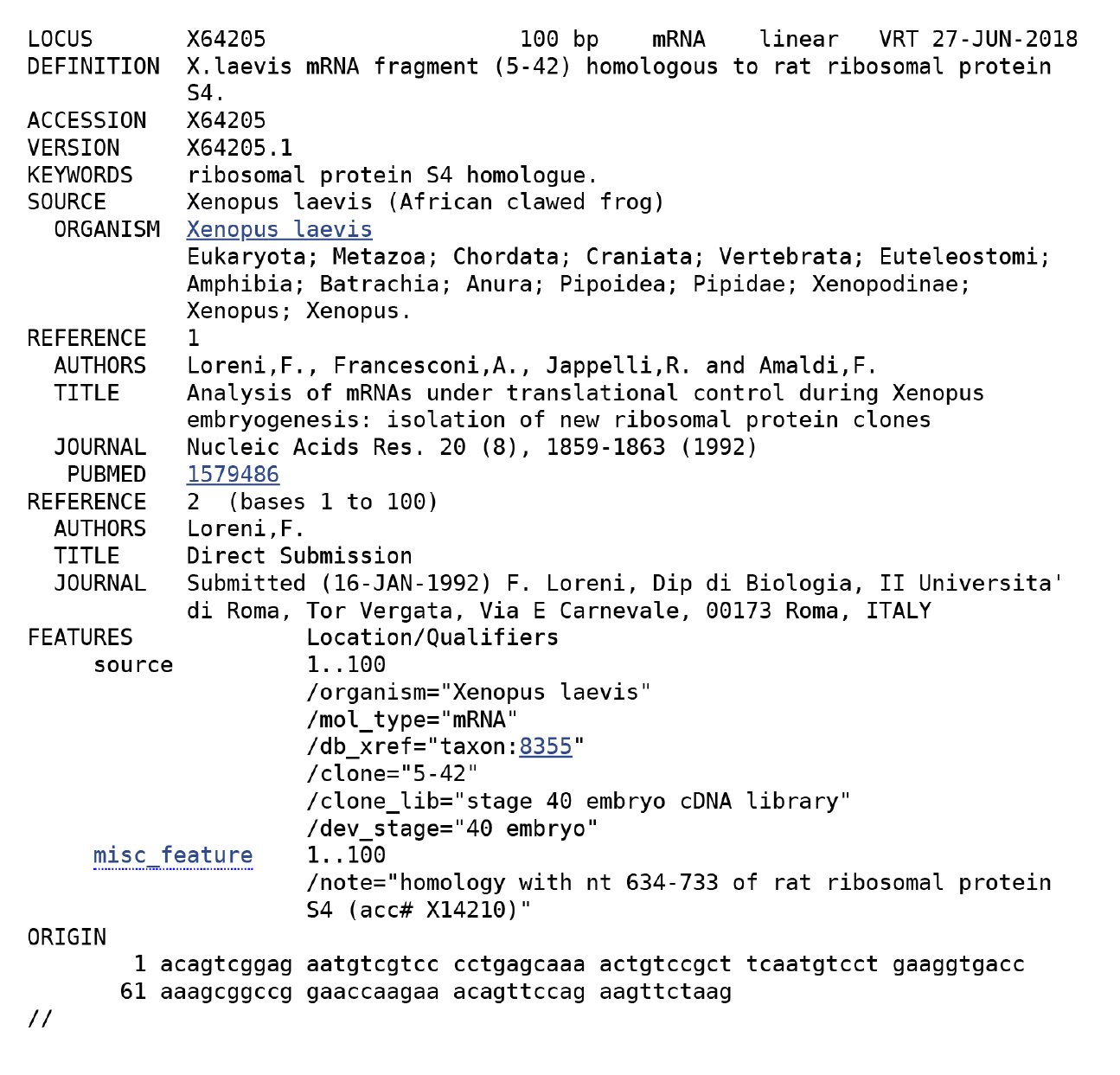}
\par\end{centering}
\caption{\label{fig:sample-data-of} Sample record of a sequence submission
from GenBank. The ACCESSION field is the unique identifier of a dataset.}
\end{figure}

GenBank is a genetic sequence database that contains an annotated
collection of all publicly available nucleotide sequences for almost
420,000 formally described species \citep{Sayers2019}. The information
about a nucleotide sequence in GenBank is organized as a record consisting
of many data elements (fields) and stored in a flat-file (see Fig.
\ref{fig:sample-data-of}). The ACCESSION field contains the unique
identifier of the dataset. The last citation in the REFERENCE field
contains the information about the submitter, including the author
list and the date in which the dataset was introduced (e.g., ``16-JAN-1992''
in Fig. \ref{fig:sample-data-of}). We obtained a snapshot of the
GenBank database (version 230) with 212,260,377 gene sequences. We
remove those sequences without submission date. This left us with
77,149,105 sequences.

The National Institutes of Health (NIH) provided us with number of
visits to a sequence's landing page for the top 1000 Nucleotide sequences
during the month of September 2012. We use these visits as a measure
of real \emph{usage.}

\subsubsection{Figshare}

Figshare is a multidisciplinary, open access research repository where
researchers can deposit and share their research output. Figshare
allows users to upload various formats of research output, including
figures, datasets, and other media \citep{thelwall2016figshare}.
In order to encourage data sharing, all research data made publicly
available has unlimited storage space and is allocated a DOI. This
DOI is used by scientists to cite Figshare resources using traditional
citation methods. Figshare makes the research data publicly and permanently
available which mitigates the resource decay problem and improves
research reproducibility \citep{zeng2019deadscience}. A Figshare
DOI contains the string 'figshare' (e.g., '10.6084/m9.figshare.6741260').
We can leverage this feature to determine whether a publication cites
Figshare resources by detecting Figshare-generated DOIs.

Figshare also keeps track of dataset accesses, such as page views
and dataset downloads. We get the downloads statistics of Figshare
DOI from the Figshare Stats API\footnote{https://docs.figshare.com/\#stats}
and use it as a measure of real \textit{usage}.

\subsection{Construction of citation network}

\begin{figure}
\begin{centering}
\includegraphics[width=1\columnwidth]{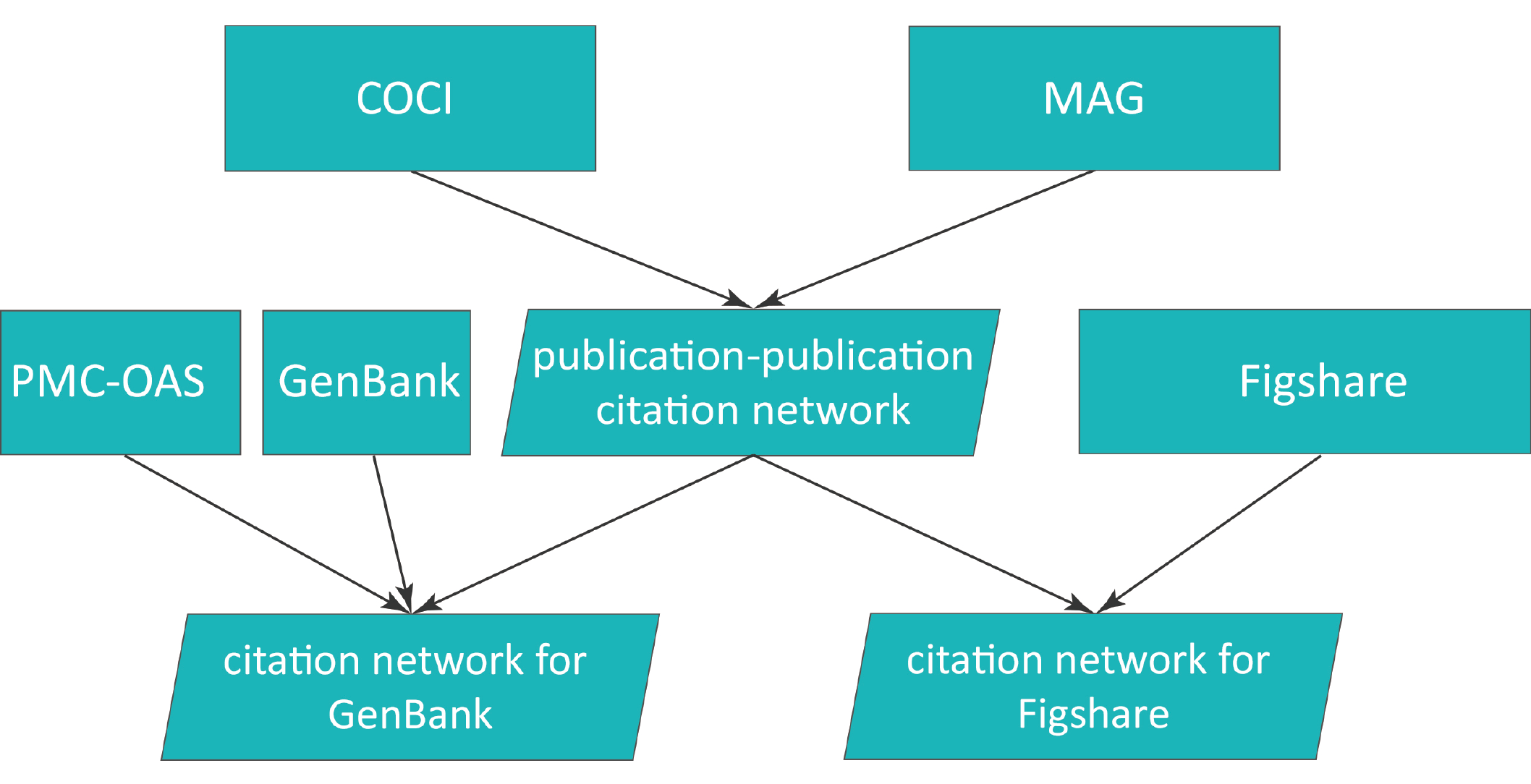}
\par\end{centering}
\caption{\label{fig:Network-construction-workflow}Citation network construction
workflow}
\end{figure}

The citation networks in this paper consist of two types of nodes
and two types of edges. The node is represented by the paper and the
dataset and the edge is represented by the citation between two nodes.
Concretely, papers cite each other to form paper-paper edges as datasets
can only be cited by papers which are represented by the paper-dataset
edges. As shown in the construction workflow (Fig. \ref{fig:Network-construction-workflow}),
we build the paper-paper citation network using COCI and MAG and build
two separate paper-dataset edge sets using GenBank and Figshare. Then
we integrate the paper-dataset edge sets into the paper-paper citation
network to form two complete citation networks. The construction workflow
is illustrated in Figure \ref{fig:Network-construction-workflow}.

\subsubsection{Paper-paper citation network}

As our proposed method takes the publication year into account, we
need to remove those edges without publication year. The DOI-DOI citation
in COCI dataset provides us a skeleton of the citation network. Even
though COCI comes with a field named timespan which originally refers
to the publication time span between the citing DOI and cited DOI,
this timespan leads to different results depending on the time granularity
(i.e., year-only vs year+month). As explained above, we have to complement
the COCI dataset with MAG to complete the year of publication of cited
and citing articles. By joining the COCI and MAG, we build a large
paper-paper citation network consisting of 45,180,243 nodes and 443,055,788
edges. As the usage data of GenBank only covers 2012 (see Genbank
above), we pruned the network to remove papers and datasets published
after 2012. For Figshare, we use all the nodes and edges. The statistics
of the networks described above are listed in Table \ref{tab:Statistics-of-network}.

\begin{table}
\caption{\label{tab:Statistics-of-network}Statistics of article citation networks}

\centering{}%
\begin{tabular}{cccc}
\hline 
Sources & Filter & Number of Nodes & Number of Edges\tabularnewline
\hline 
COCI (original) & N/A & 46,534,424 & 449,840,585\tabularnewline
COCI (filtered) & contains publication year & 21,689,394 & 203,884,791\tabularnewline
COCI \& MAG (for Figshare) & contains publication year & 45,180,243 & 443,055,788\tabularnewline
COCI \& MAG (for GenBank) & publication year \ensuremath{\le} 2012 & 30,304,869 & 212,410,743\tabularnewline
\hline 
\end{tabular}
\end{table}

\subsubsection{Paper-dataset citation network}
\begin{enumerate}
\item \textbf{Constructing paper-dataset citations for GenBank. }As previously
described by \citet{Sayers2019}, authors should use GenBank accession
number with the version suffix as identifier to cite a GenBank data.
The accession number is usually mentioned in the body of the manuscript.
This practice enables us to extract GenBank dataset mention from the
PMC-OAS dataset to build the paper-dataset citation network. 
\begin{enumerate}
\item \emph{Parsing XML file to extract full-text}\textbf{. }We first parse
the XML files using XPath expressions to extract the PMID and the
full-text. We get 2,174,782 articles with PMID and full-text from
PMC-OAS dataset.
\item \emph{Matching the accession number to build paper-dataset citations}.
According to the GenBank Accession Prefix Format, an accession number
is composed of a fixed-number of letters plus a fixed-number of numerals
(e.g., 1 letter + 5 numerals, 2 letters + 6 numerals). Based on the
format, we composed a regular expression to match individual mentions
of accession numbers in the full-text. There are two kinds of accession
number mentions: accession number only (e.g., U00096) and range of
accession number (e.g., KK037225-KK037232). For the second kind, we
expanded the range to recover all the omitted accession numbers.
\end{enumerate}
\end{enumerate}
\begin{enumerate}[resume]
\item \textbf{Extracting paper-dataset citation for Figshare. }As there
is a string 'figshare' in a Figshare DOI, we can use a regular expression
to search for them in COCI. 
\begin{enumerate}
\item \textit{Identifying Figshare DOI in the set of cited DOI. }In this
step, we extracted 918 Figshare DOIs as dataset candidates.
\item \textit{Filtering by resource type. }Not all the Figshare DOIs are
datasets because Figshare supports many kinds of resources. We use
the Figshare Article API\footnote{https://docs.figshare.com/\#public\_article}
to get the meta-data of a DOI. In the meta-data, there is a field
indicating the resource type (type code 3 is dataset). After filtering
the candidates by resource type, we get 355 datasets.
\end{enumerate}
\end{enumerate}

\subsubsection{Visualization of citation network}

\begin{figure}
\begin{centering}
\includegraphics[width=1\textwidth]{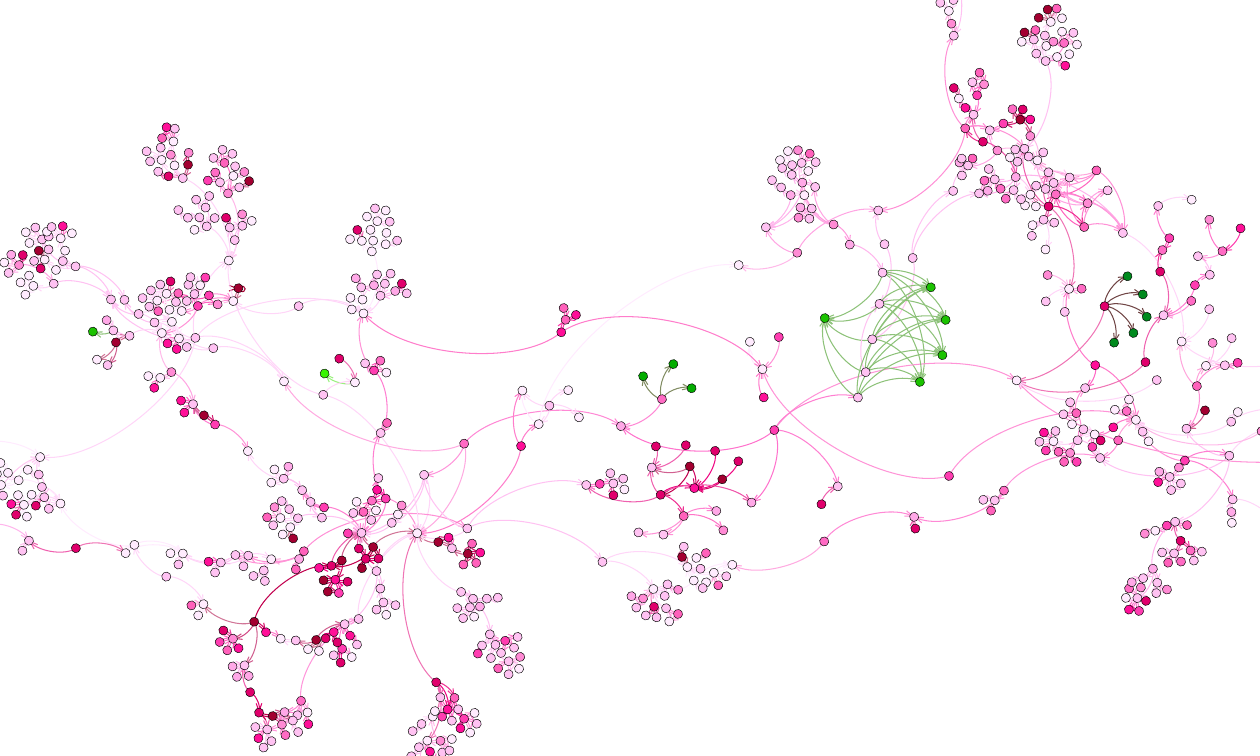}
\par\end{centering}
\caption{\label{fig:network-of-sample} Visualization of publication--dataset
citation network. Papers appear as red nodes and Datasets are green.
The lighter the color, the older the resource.}
\end{figure}

For the purpose of exploring this network, we sample about one thousand
nodes and 1.5 thousand edges from the network. We observe four patterns
of papers citing datasets (a dataset cannot cite anything): one-to-one,
one-to-many, many-to-one and many-to-many (Fig. \ref{fig:network-of-sample}).

\subsection{Network models for scientific artifacts}

\subsubsection{\textsc{NetworkFlow}}

We adapt the network model proposed in \citet{walker2007ranking}.
This method, which we call \textsc{NetworkFlow} here, is inspired
by \textsc{PageRank} and addresses the issue that citation networks
are always directed back in time. In this model, each vertex in the
graph is a publication. The goal of this method is to propagate a
vertex's impact through its citations.

This method simulates a set of researchers traversing publications
through the citation network. It ranks publications by estimating
the average path length taken by researchers who traverse the network
considering the age of resources and a stopping criterion. Mathematically,
it defines a traffic function $T_{i}(\tau_{\text{pub}},\alpha)$ for
each paper. A starting paper is selected randomly with a probability
that exponentially decays in time with a \emph{decay} parameter $\tau_{\text{paper}}$.
Each occasion the researcher moves, it can stop with a probability
$\alpha$ or continue the search through the citation network with
a probability $1-\alpha$. The predicted traffic $T_{i}$ is proportional
to the rate at which the paper is accessed. The concrete functional
form is as follows. The probability of starting at the $i$th node
is 
\begin{equation}
\rho_{i}\propto\exp\left(-\frac{\text{age}_{i}}{\tau_{\text{pub}}}\right)\label{eq:starting-probability}
\end{equation}

Then,\textsc{ }the method defines a transition matrix from the citation
network as follows 
\begin{equation}
W_{ij}=\begin{cases}
\frac{1}{k_{j}^{\text{out}}} & \text{if \ensuremath{j} cites \ensuremath{i}}\\
0 & \text{o.w.}
\end{cases}\label{eq:transition-probability}
\end{equation}
where $k_{j}^{\text{out}}$ is the out-degree of the $j$th paper.

The average path length to all papers in the network starting from
a paper sampled from the distribution $\rho$ is defined as 
\begin{equation}
T=I\cdot\rho+(1-\alpha)W\cdot\rho+(1-\alpha)^{2}W^{2}\rho+\cdots\label{eq:average-path-length}
\end{equation}

The parameters $\tau_{\text{paper}}$ and $\alpha$ are found by cross
validation by predicting real traffic. In practice, this series can
be solved iteratively by computing the difference between $T_{t+1}$
and $T_{t}$. For this and all algorithms used in this work, we stop
the iterations when the total absolute difference in rank between
two consecutive iterations is less than $10^{-2}$ which typically
required around 30 iterations.

\subsubsection{\textsc{DataRank}}

In this article, we extend\textsc{ NetworkFlow} to accommodate different
kinds of nodes. The extension considers that the probability of starting
at any single node should depend on whether the node is a publication
or a dataset. This is, publications and datasets may vary in their
relevance period. We call this new algorithm \textsc{DataRank}. Mathematically,
we redefine the starting probability in Eq. \ref{eq:starting-probability}
of the $i$th node as 
\begin{equation}
\rho_{i}^{\text{DataRank}}=\begin{cases}
\exp\left(-\frac{\text{age}_{i}}{\tau_{\text{pub}}}\right) & \text{if \ensuremath{i} is a publication}\\
\exp\left(-\frac{\text{age}_{i}}{\tau_{\text{dataset}}}\right) & \text{\text{if \ensuremath{i} is a dataset}}
\end{cases}\label{eq:data-rank-starting}
\end{equation}

\begin{figure}
\begin{centering}
\includegraphics[width=0.95\textwidth]{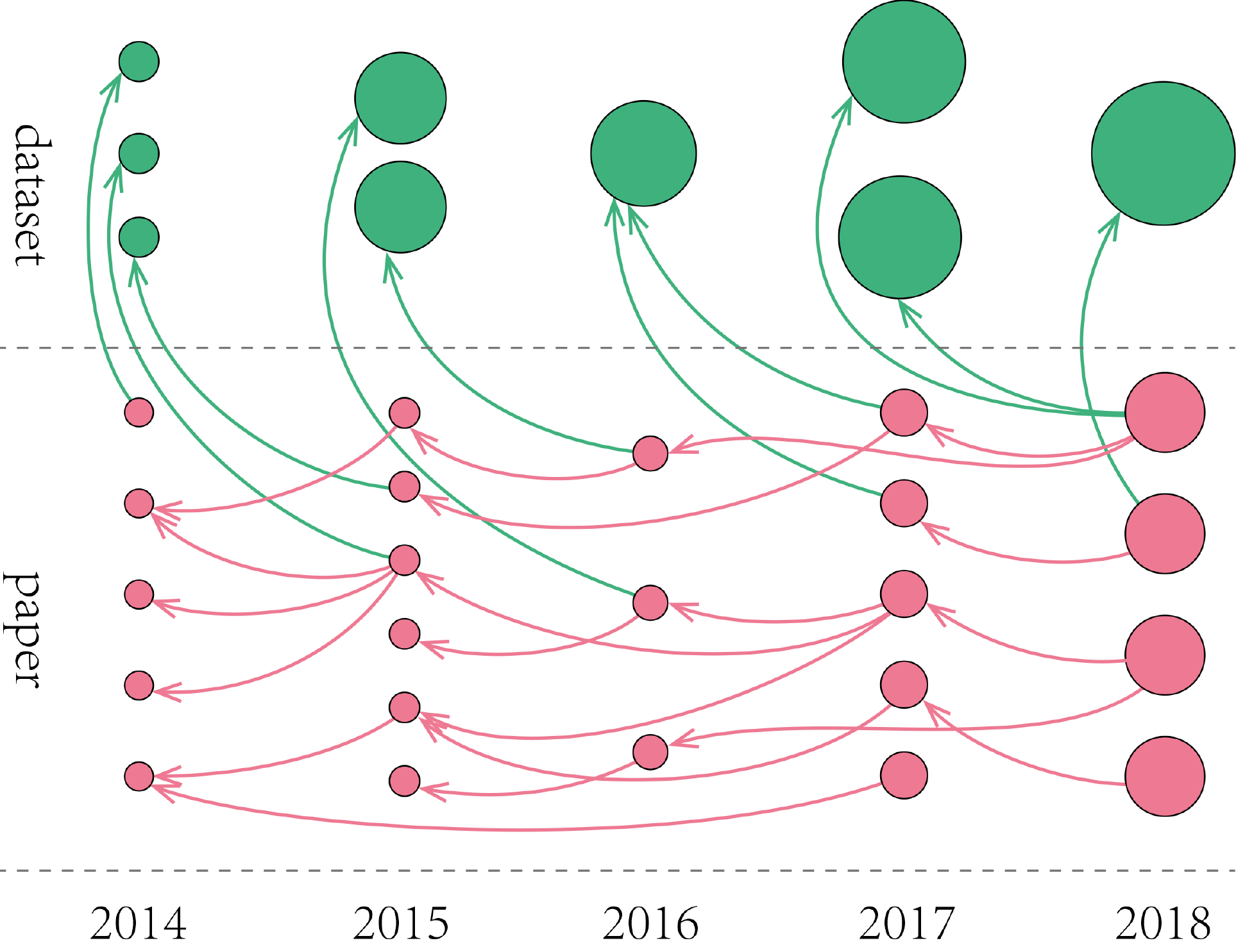}
\par\end{centering}
\caption{\label{fig:Initialize-the-value}Nodes after initialization. The size
of the node represents its value, different color means different
node type: red nodes are papers and green nodes are datasets.}
\end{figure}

This initialization process is depicted in Figure \ref{fig:Initialize-the-value}.
Here, datasets have a smaller decay than papers. The size of the bubble
indicates the initial flow as defined by Eq. \ref{eq:data-rank-starting}.
After initializing, we can easily find that nodes of the same type
and same age have the same value, and that the younger a node is,
the bigger its value.

Now, we estimate the traffic in a similar fashion but now the traffic
of a node $T_{i}^{\text{DataRank}}(\tau_{\text{pub}},\tau_{\text{dataset}},\alpha)$
depends on three parameters that should be found by cross-validation
with the rest of the components of the method remaining the same (e.g.,
\eqref{transition-probability} and \eqref{average-path-length}).

\subsubsection{\textsc{DataRank-FB}}

In \textsc{DataRank}, each time the walker moves, there are two options:
to stop with a probability $\alpha$ or to continue the search through
the reference list with a probability $1-\alpha$. However, there
may exist a third choice: to continue the search through papers who
cite the current paper. In other words, the researcher can move in
two directions: forwards and backwards. We call this modified method
\textsc{DataRank-FB. }In this method, one may stop with a probability
$\alpha-\beta$, continue the search forward with a probability $1-\alpha$,
and backward with a probability $\beta$. To keep them within the
unity simplex, the parameters must satisfy $\alpha>0$, $\beta>0$,
and $\alpha>\beta$,

Then\textbf{\textsc{, }}we define another transition matrix from the
citation network as follows 
\begin{equation}
M_{ij}=\begin{cases}
\frac{1}{k_{j}^{\text{in}}} & \text{if \ensuremath{j} cited by \ensuremath{i}}\\
0 & \text{o.w.}
\end{cases}\label{eq:transition-probability-1}
\end{equation}
where $k_{j}^{\text{in}}$ is the number of papers that cite $j$.
We update the average path length to all papers in the network starting
from $\rho$ as

\begin{equation}
T=I\cdot\rho+(1-\alpha)W\cdot\rho+\beta M\cdot\rho+(1-\alpha)^{2}W^{2}\rho+\beta^{2}M^{2}\rho+\cdots\label{eq:average-path-length-1}
\end{equation}

\subsection{Other network models}

\subsubsection{\textsc{PageRank}}

\textsc{PageRank} is a well-known and widely used webpage ranking
algorithm proposed by Google \citep{page1999pagerank}. It uses the
topological structure of the web to determine the importance of a
webpage, independently of its content \citep{bianchini2005inside}.
First, it builds a network of the web through the link between webpages.
Second, each webpage is assigned a random value, which is then updated
based on the link relationships---an iterative process that will
eventually converge to a stationary value.

The mathematical formulation of \textsc{PageRank} is

\begin{equation}
PR(p_{i})=\frac{1-d}{N}+d\underset{p_{j}\in M(p_{i})}{\sum\frac{PR(p_{j})}{L(p_{j})}},\label{eq:pagerank-algorithm}
\end{equation}
where $p_{1},p_{2},\cdots,p_{N}$ are the webpages whose importance
need to be calculated, $M(p_{i})$ is the set of webpages that has
a link to page $p_{i}$, $L(p_{j})$ is the number of outbound links
on webpage $p_{j}$, and $N$ is the total number of webpages. The
parameter $d$ is a dampening factor which ranges from 0 to 1 and
is usually set to 0.85 \citep{brin1998anatomy,page1999pagerank}.

\subsubsection{Modified \textsc{PageRank}}

Considering that we have two types of resources in the network, we
modify the standard \textsc{PageRank} to allow a different damping
factor for publication and for dataset. This amounts to modifying
the update equation to

\begin{equation}
PR(p_{i})=\frac{1-d_{data}}{N_{data}}+\frac{1-d_{pub}}{N_{pub}}+d_{data}\underset{p_{j}\in M^{data}(p_{i})}{\sum\frac{PR(p_{j})}{L(p_{j})}}+d_{pub}\underset{p_{k}\in M^{pub}(p_{i})}{\sum\frac{PR(p_{k})}{L(p_{k})}},\label{eq:modified-pagerank}
\end{equation}
where $p_{1},p_{2},\cdots,p_{N}$ are still the nodes whose importance
need to be calculated, $M^{data}(p_{i})$ is the set of datasets that
have a link to page $p_{i}$, $M^{pub}(p_{i})$ is the set of papers
that has a link to page $p_{i}$, $L(p_{j})$ is the number of outbound
links on webpage $p_{j}$, and $N_{data},N_{pub}$ are the total size
of datasets and paper collections, respectively. The parameter $d_{data}$
is a damping factor for datasets and $d_{pub}$ is a damping factor
for paper sets.

\section{Results}

We aim at finding whether the estimation of the rank of a dataset
based on citation data is related to a real-world measure of relevance
such as page views or downloads. We propose a method for estimating
rankings that we call \textbf{\textsc{DataRank,}} which considers
differences in citation dynamics for publications and datasets. We
also propose some variants to this method, and compare all of them
to standard ranking algorithms. We use the data of visits of GenBank
datasets and downloads of Figshare datasets as measure of real usage.
Thus, we will investigate which of the methods work best for ranking
them.

\subsection{Properties of the citation networks}

\begin{figure}
\begin{centering}
\includegraphics[width=0.8\textwidth]{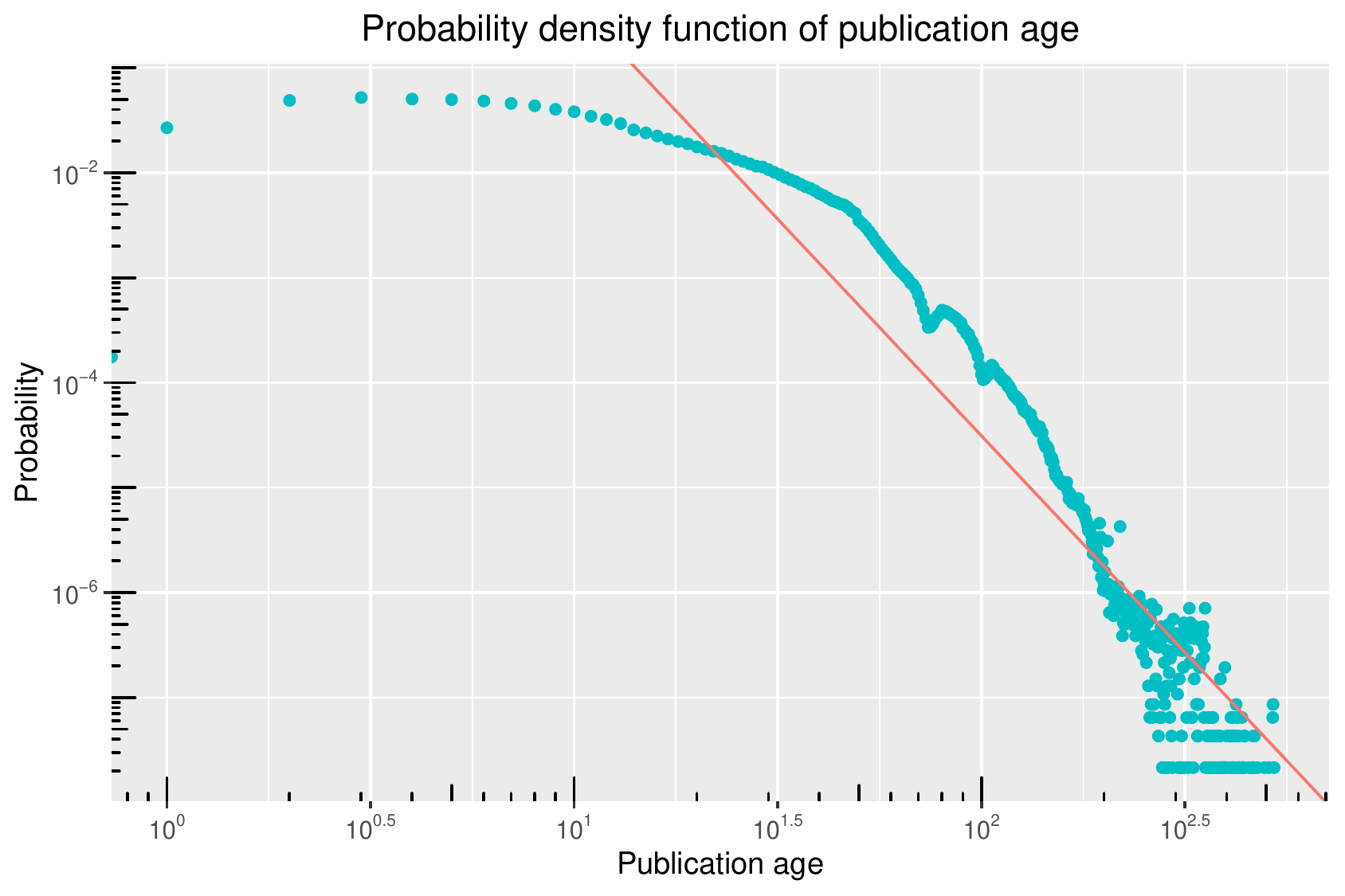}
\par\end{centering}
\begin{centering}
\includegraphics[width=0.8\textwidth]{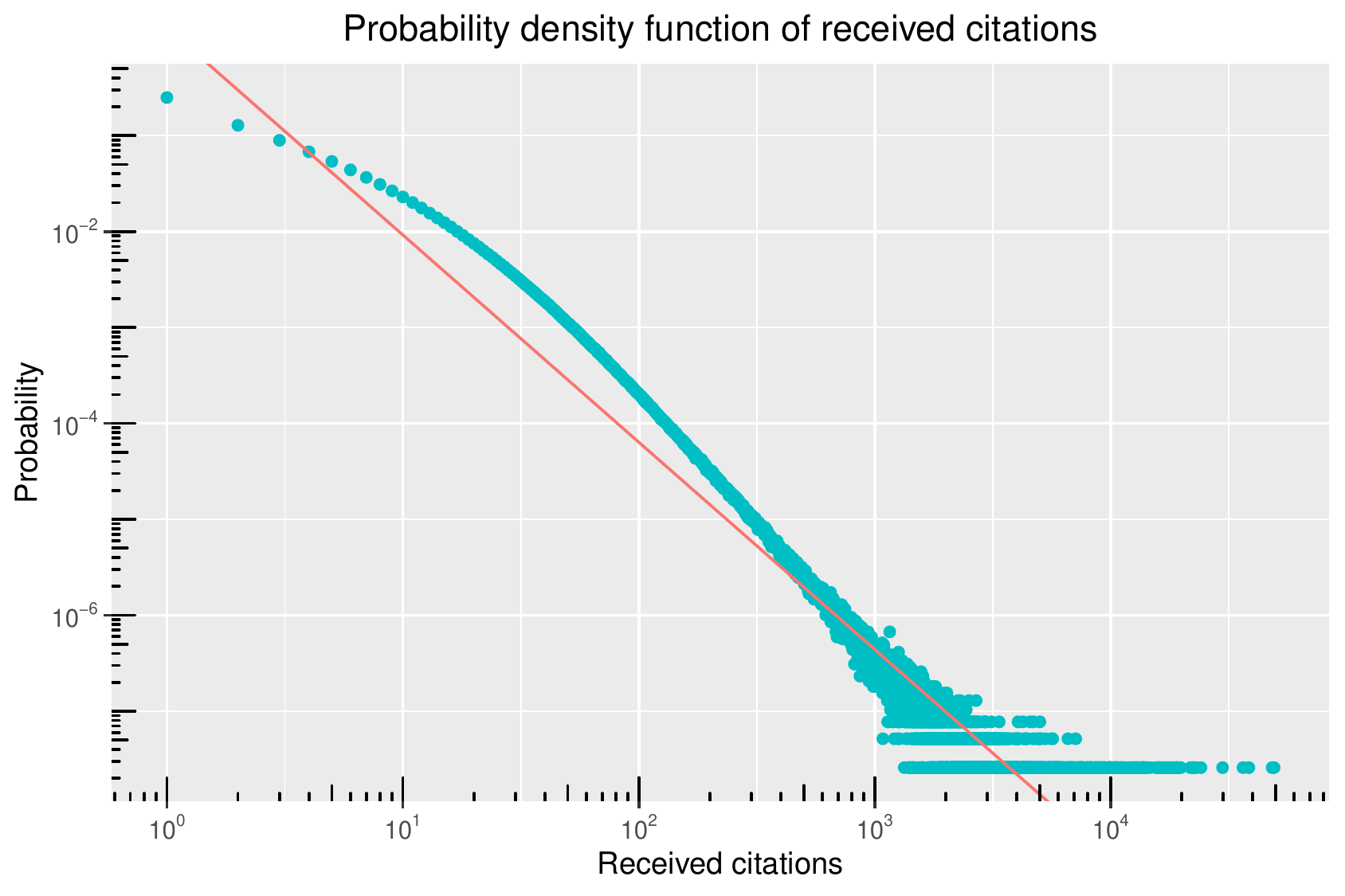}
\par\end{centering}
\caption{\label{fig:paper_network_age_citation_dist}Probability density of
age and citations. Power laws describing the growth of both aspects
of the network have parameters $k_{\text{age}}\approx-4.1$ and $k_{\text{citations}}\approx-2.16$.}
\end{figure}

We first examined some statistics of the paper--paper citation network
and the paper--dataset citation network. Specifically, we examined
dataset age because it determines the initial rank of a node and we
examine the citations because they control how ranks diffuse through
the network. We modeled publication age with respect to 2019 as a
power law distribution $p(a)\propto a^{-k}$ and we found the best
parameter to be $k\approx-4.1$ (SE=$0.07$, $p<0.001$, Fig. \ref{fig:paper_network_age_citation_dist}
top panel). Similarly, the citation count can be modeled by a power
law distribution $p(d)\propto d^{-k}$ with parameter $k\approx-2.16062$
(SE=$0.012$, $p<0.001$, Fig. \ref{fig:paper_network_age_citation_dist}
bottom panel). While both distributions can be well described by power
laws, the age distribution had some out-of-trend dynamics for small
age values because the number of publications is not growing as fast
as the power law would predict. The networks thus is expanding fast
in nodes (e.g., age) and highly skewed for citations.

\begin{figure}
\begin{centering}
\includegraphics[width=0.8\textwidth]{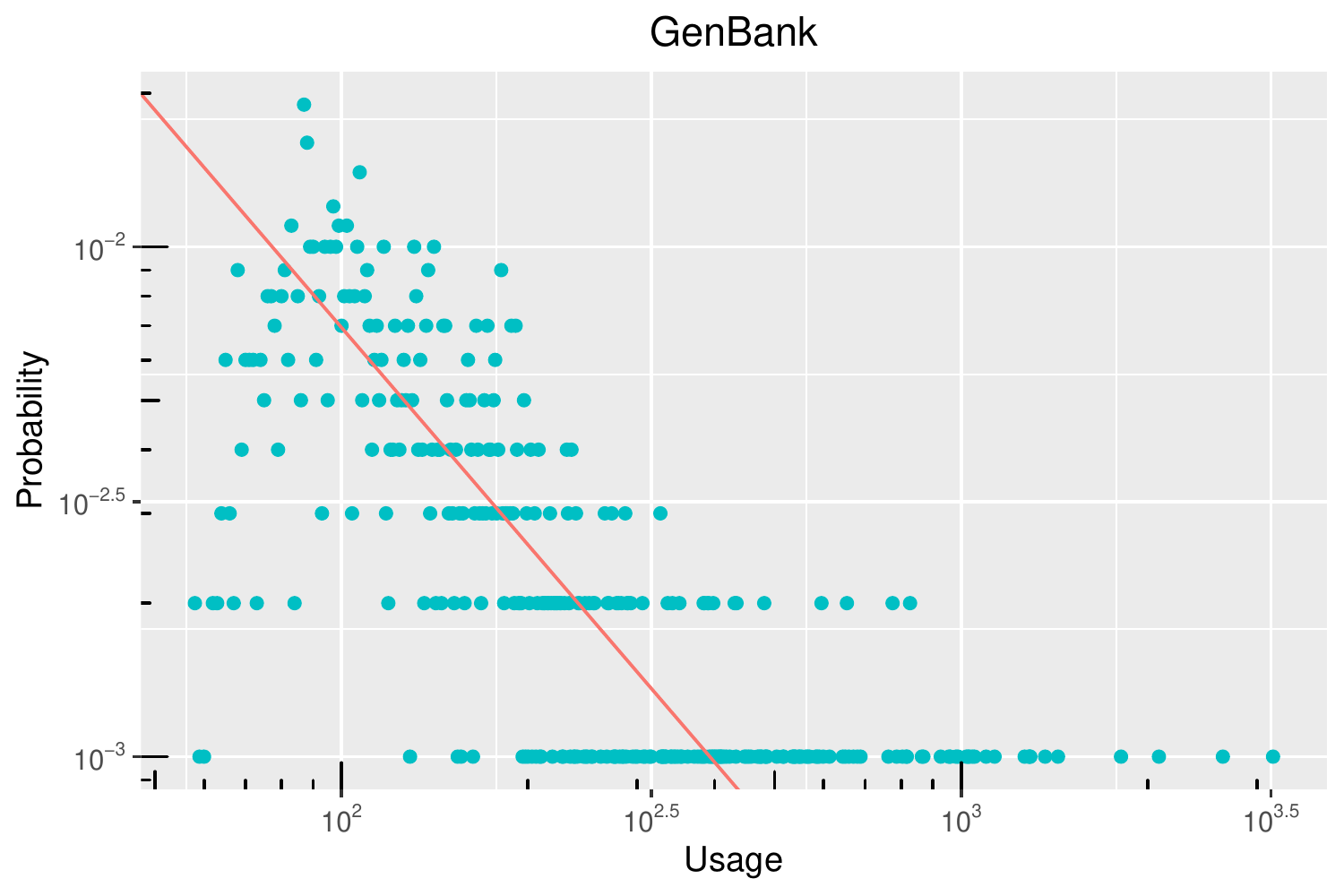}
\par\end{centering}
\begin{centering}
\includegraphics[width=0.8\textwidth]{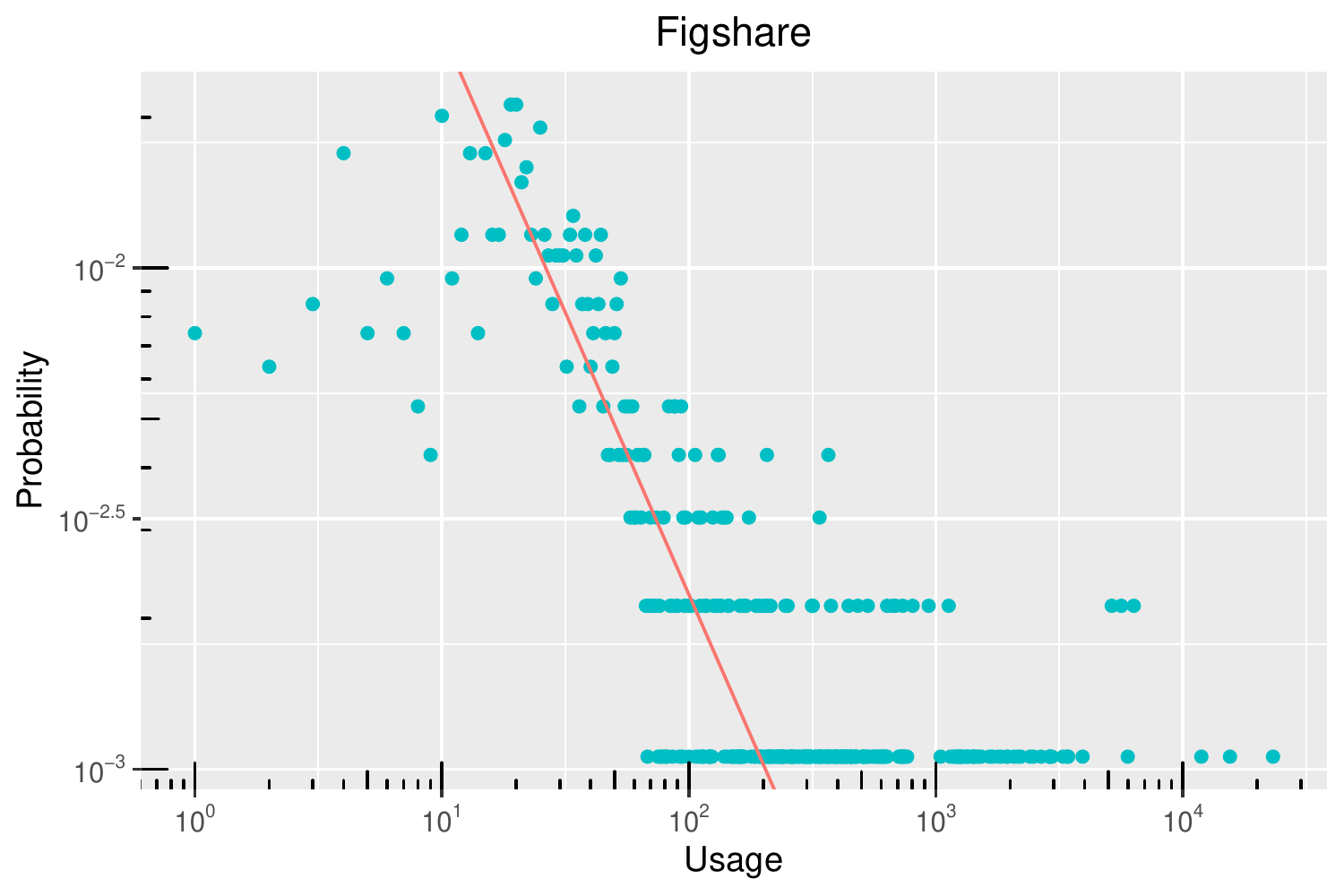}
\par\end{centering}
\caption{\label{fig:pdf-of-figshare}Probability density function of usage
(website visits for Genbank and downloads for Figshare)}
\end{figure}

We then wanted to examine differences in usage between Genbank and
Figshare. We plotted the estimated probability density function of
this usage for both datasets as a log-log plot (Fig. \ref{fig:pdf-of-figshare}).
While the scale of usage is significantly different (i.e., overall
GenBank is more used than Figshare), there seems to be a long-tail
power law relationship in usage. The GeneBank dataset had a larger
but not significantly different scale parameter than Figshare ($k\approx-1.416$,
SE=0.2699, in GenBank and $k\approx-1.125$, SE=0.0907, in Figshare,
for $p(u)\propto u^{-k}$, two-sample t-test $t(220)=-1.05$, $p=0.29$).
There is a significant bias for downloads below a certain threshold,
where smaller number of downloads are less frequent than expected
by the power laws. However, both datasets show similar patterns suggesting
a common mechanism driving the dataset usage behavior.

\subsection{Prediction of real usage}

One of the real tests of whether the methods work is to predict how
they are related to real usage data. For each algorithm and set of
parameters, we estimated the rank of networks' nodes. We then correlated
these ranks with real usage (i.e., web visits for GenBank and downloads
for Figshare). We now describe the best performance after these parameter
search.

\begin{figure}
\begin{centering}
\includegraphics[width=1\textwidth]{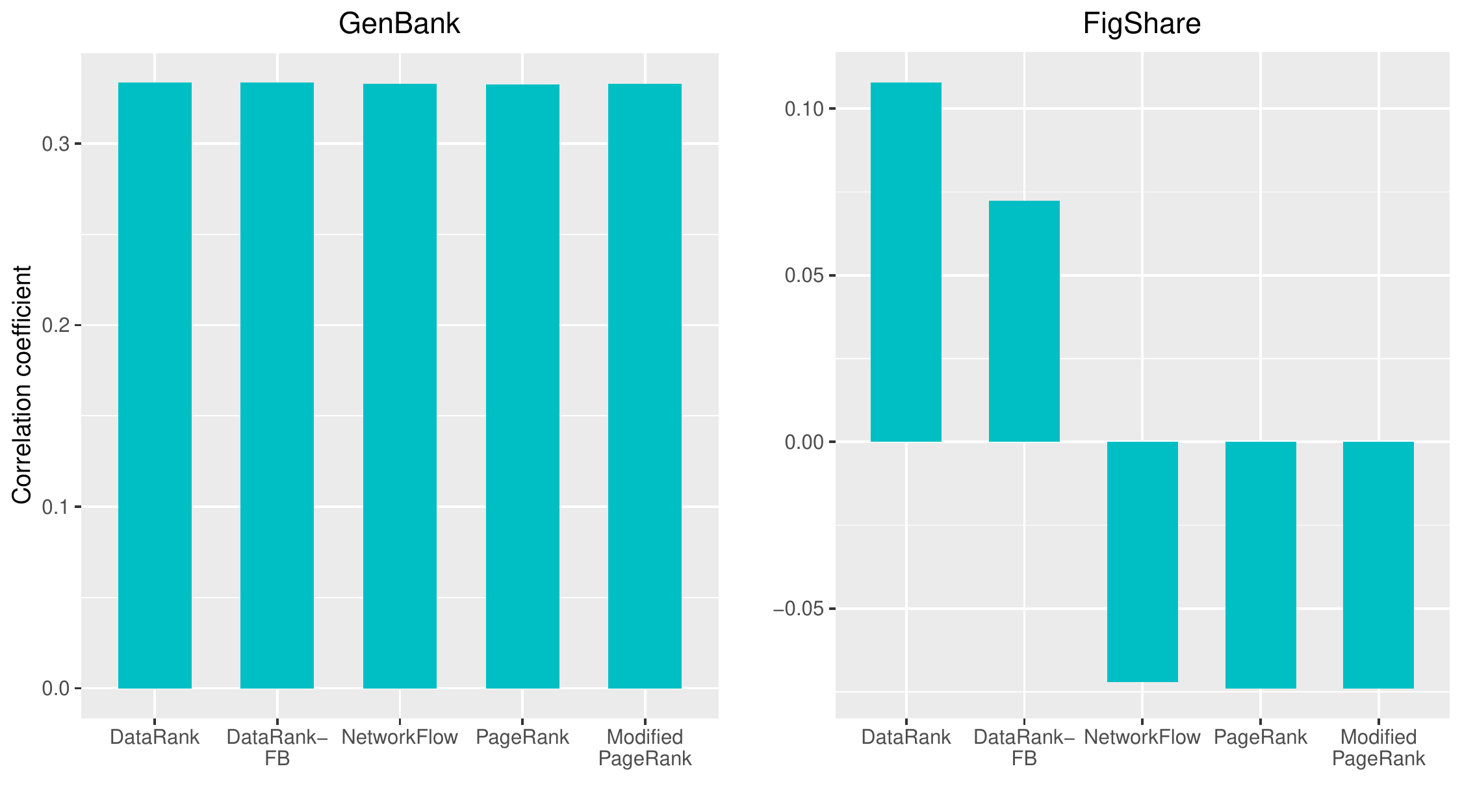}
\par\end{centering}
\caption{\label{fig:result-correlation}Model comparison in correlation coefficient.
\textsc{DataRank} is able to predict better the real usage of datasets
compared to other variants.}
\end{figure}

We performed a grid search with publication decay year $\tau_{\text{pub}}\in\left\{ 1,5,10,20,30,50,70,100\right\} $,
the dataset decay year $\tau_{\text{dataset}}\in\left\{ 1,5,10,20,30,50,70,100\right\} $,
and alpha in $\alpha\in\left\{ 0.05,0.15\right\} $. The best model
of \textsc{DataRank} achieved a correlation of 0.3336 in Genbank which
is slightly better than \textsc{DataRank-FB} 0.3335, \textsc{NetworkFlow}
0.3327, \textsc{PageRank} 0.3324, and Modified \textsc{PageRank} 0.3327
(Fig. \ref{fig:result-correlation}). The best model of \textsc{DataRank}
achieves a correlation of 0.1078 in Figshare, which is substantially
better than \textsc{DataRank-FB} with 0.0723. The best absolute correlations
by all the other methods only achieved negative correlations (\textsc{NetworkFlow}\textbf{\textsc{=}}$-0.072$,
\textsc{PageRank}=$-0.073$, and Modified \textsc{PageRank}=$-0.073$)---we
selected models based on absolute correlation because it would produce
the best predictive performance. Taken together, \textsc{DataRank}
has both better absolute correlation performance and highest correlation,
suggesting a superior ability to predict real usage.

\subsection{Interpretation of best-fit model}

We wanted to explore how \textsc{DataRank} changes in performance
across the different parameters. This exploration allowed us to understand
how the parameters tell us something about the underlying characteristics
of the citation networks.

\begin{figure}
\begin{centering}
\includegraphics[width=0.8\textwidth]{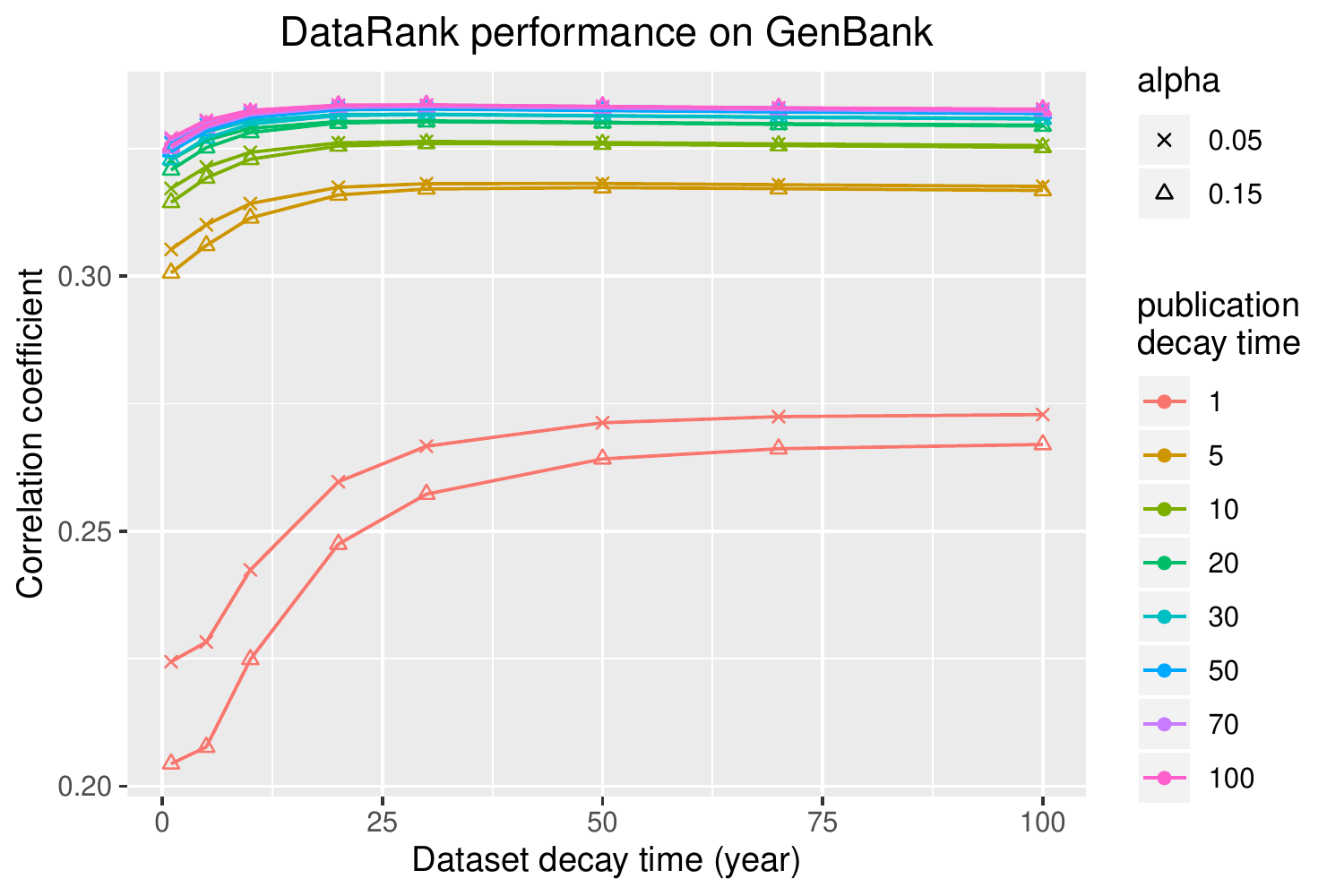}
\par\end{centering}
\begin{centering}
\includegraphics[width=0.8\textwidth]{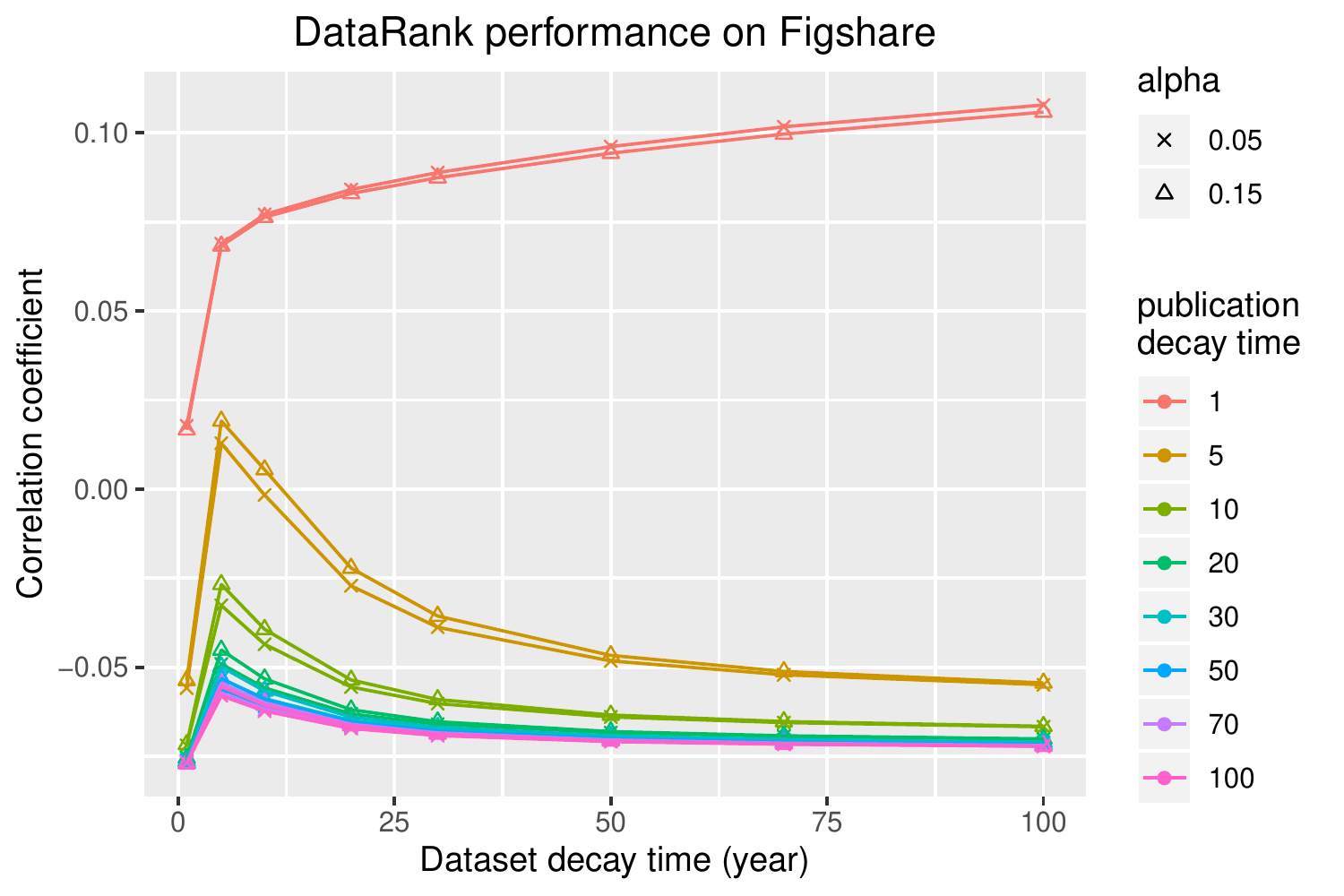}
\par\end{centering}
\caption{\label{fig:Parameter-search}Model performance during grid search
with publication decay time (years) $\tau_{\text{pub}}\in\left\{ 1,5,10,20,30,50,70,100\right\} $,
the dataset decay time (years) $\tau_{\text{dataset}}\in\left\{ 1,5,10,20,30,50,70,100\right\} $,
and alpha in $\alpha\in\left\{ 0.05,0.15\right\} $}

\end{figure}

The results across all parameters are shown in Figure \ref{fig:Parameter-search}
and we now describe general trends. For experiments on GenBank dataset,
we found the \textsc{DataRank} model reaches its best with $\tau_{\text{pub}}=100$
and $\tau_{\text{dataset}}=30$ (Fig. \ref{fig:result-correlation}
top panel). However, for dataset decay, the performance reaches a
plateau after around 20 years and it reaches its peak at 30 years.
After 30 years, performance goes down slowly. In terms of publication
decay, the performance increases significantly before 20 years. After
20 years, the performance enters into a steady but marginal increase.
In all the cases, the alpha at 0.05 is better than 0.15. For Figshare
dataset, dataset decay has divergent patterns: for small publication
decays, dataset decay increases the performance. For large publication
decays, dataset decay decreases the performance. For publication decay,
more specifically, we observed an opposite trend compared to GenBank:
the performance goes down rapidly as the publication decay increases
(Fig. \ref{fig:result-correlation} bottom panel). However, similar
to GenBank, it losses momentum after 20 years but still decreases
steadily.

One explanation for the differences between the best parameters for
Genbank and Figshare is that dataset age and citation distribution
have different patterns. We perform an analysis to confirm this hypothesis.
Indeed, we found that Figshare datasets are significantly younger
than Genbank datasets (bootstrapped difference between average age
= $-4.80$ years, SE = 0.24, $p<0.001$). We also found that Figshare
dataset citations are more uniform than Genbank dataset citations
although not significantly different (bootstrapped difference in dataset
citation kurtosis = -44.34, SE = 48.74, $p=0.25$). The \textsc{DataRank}
algorithm uses large values of dataset decay to propagate publication
citations over a longer time and small values of data decay to steer
those citations towards a concentrated set of top cited datasets.
This is the case of GenBank dataset dynamics. Figshare, comparatively,
obeys opposite age and citation dynamics which \textsc{DataRank} attempts
to accommodate by using small values of publication decay and large
values of dataset decays. Concretely, the best publication decay for
Genbank is 100 and for Figshare is 1, and the best dataset decay for
Genbank is 30 and for Figshare is 100. Taken together, \textsc{DataRank}'s
parameters offer inferential interpretation of the citation network
dynamics, which can help us understand how they are distributed in
time and credit spaces.

\section{Discussion}

The goal of this article is to better evaluate the importance of datasets
through article citation network analysis. Compared with the mature
citation mechanisms of articles, referencing datasets is still in
its infancy. Acknowledging the long time the practice of citing datasets
will take to be adopted, our research aims at recovering the true
importance of datasets even if their citations are biased compared
to publications.

Scholars disagree on how to give credit to research outcomes. Regardless
of how disputed citations are as a measure of credit \citep[see][]{moed1985use,diamond1986citation,martin1996use,seglen1997impact,wallin2005bibliometric},
they complement other measures that are harder to quantify such as
peer review assessment \citep{meho2007impact,piwowar2007sharing}
or real usage such as downloads \citep{belter2014measuring}. Citations,
however, are rarely used for datasets \citep{altman2013evolution},
giving these important research outcomes less credit that they might
deserve. Our proposal aims at solving some of these issues by constructing
a network flow that is able to successfully predict real usage better
than other methods. While citations are not a perfect measure of credit,
having a method that can at least attempt to predict usage is advantageous.

Previous research has examined ways of normalizing citations by time,
field, and quality. This relates to our \textsc{DataRank} algorithm
in that we are trying to normalize the citations by year and artifact
type. Similar work has been done in patent citations: \citet{hall2001nber}
attempt to eliminate the effects caused by year, field and year-field
interaction through dividing the number of citation received by a
certain patent by the corresponding year-field average number of citation
of patents of each cohort in each field. \citet{yang2015using} used
patent citation networks to propagate citations and evaluate patent
value. Other researchers have gone beyond time and field by also attempting
to control by quality of the resource. Clarivate Analytics editors
consider other normalizing factors such as publishing standards, editorial
content and citation information when creating the Data Citation Index
(DCI) database \citep{reuters2012repository}. However, a control
like this requires significantly more manual curation.

We found that the practice of following who cites an artifact (i.e.,
backward propagation) seemed to be not as important compared to following
a cited artifact (i.e., forward propagation). This is, the more specialized
\textsc{DataRank-FB} model (see Methods) did not produce higher predictability.
This lack of improvement could suggest that when scientists traverse
the network of citations, they seem to only follow one direction of
the graph (i.e., the reference list). This could be perhaps a limitation
of the tools available to scientists to explore the citation network
or because backward propagation changes constantly year after year.
Initiatives that force the creation of identifiers (DOI) for datasets,
such as the NERC Science Information Strategy Data Citation and Publication
project \citep{callaghan2012making}, might change this pattern.

We find it useful to interpret the effect of different decay times
on the performance of \textsc{DataRank}. The best-fitting parameters
show that \textsc{DataRank} attempts to model the network temporal
and topological dynamics differently for Genbank and Figshare. Because
Genbank tends to have older datasets that have more concentrated citations
compared to Figshare, the large value in publication decay time and
small value in dataset decay time produces rank distributions that
match the underlying dynamics of Genbank (Fig. \ref{fig:Parameter-search}).
Similarly, the alpha parameter, which controls the probability of
the random walk to stop, is larger for Figshare, intuitively suggesting
that due to the smaller network size and shorter temporal paths, a
higher probability of stopping should be in place (Fig. \ref{fig:Parameter-search}).
Thus, \textsc{DataRank} can help to interpret dataset citation behavior.

There are some shortcomings in our study. For GenBank and Fighsare,
we do not have a great deal of information on actual usage. There
are 1000 records but we only located 693 in the publication network
of Genbank and 355 datasets from Figshare. Also, there are significant
differences in the information, with Figshare dataset being cited
less often than GenBank. In the future, we will explore other data
repositories such as PANGEA, Animal QTLdb, and UK Data Archive, and
we will request updates about web visits to other GenBank sequences.

Open research datasets have little means for systematic attribution
even in well-resourced disciplines, such as the biomedical community,
and our proposal attempts to systematize this attribution more broadly
without requiring changes in behavior. The lack of systematic benchmarking
tools prevents good policy-making. Furthermore, it contributes to
the invisibility of digital objects and labor, an issue of serious
concern \citep{scroggins2019labor}. When digital objects are not
well described, they tend to disappear from view. They fall to the
bottom of classic search engine results when there are no mechanisms
to assign credit to datasets. This lack of indexing creates an asymmetrical
representation of some kinds of objects of science and can be an obstacle
to quality evaluation and data reuse. Thus \textsc{DataRank} can be
used to correct these asymmetrical discrepancies between articles
and datasets.

Our approach does not directly estimate true impact but predicts one
possible measure of usage. However, this prediction could be a foundational
step toward developing technical and theoretical models of impact.
Also, in the literature, \textquotedblleft impact\textquotedblright{}
is defined differently depending on the goal \citep{piwowar2013altmetrics}.
Metrics estimating impact can be defined through \textquotedblleft use\textquotedblright ,
\textquotedblleft reuse\textquotedblright , or \textquotedblleft engagement\textquotedblright ,
and have a range of proxies. For example, a download might measure
use while in another context it may only indicate viewing. Comparatively,
the edifice of citation standards in the realm of journal articles
is relatively well-established, with refinements to the interpretation
of citation behavior such as the relative value of citations \citep{stuart2017data},
the disciplinary context \citep{borgman1989bibliometrics}, and the
in-text location \citep{teplitskiy2018almost}. Nonetheless, usage
serves as a proximal indicator of influence, and a first-order approximation
of the impact of the scientific work. Recent work in developing impact
assessment tools continue development and refinement \citep{silvello2018theory}.
However, existing evaluation mechanisms often fail where there are
no direct or indirect measure for usage, which can be an important
indicator of scientific impact. Clearly, our approach serves not only
as a concrete tool to measure impact---albeit imperfectly---but
also as a vehicle to discuss why and how to measure impact on datasets.

\section{Conclusion}

Understanding how datasets are used is an important topic in science.
Datasets are becoming crucial for the reproduction of results and
the acceleration of scientific discoveries. Scientists, however, tend
to dismiss citing datasets and therefore there is no proper measurement
of how impactful datasets are. Our method uses the publication-publication
citation network to propagate the impact to the publication--dataset
citation network. Using two databases of real dataset networks, we
demonstrate how our method is able to predict actual usage more accurately
than other methods. Our results suggest that datasets have different
citation dynamics to those of publications. In sum, our study provides
a prescriptive model to understand how citations interact in publication
and dataset citation networks and gives a concrete method for producing
ranks that are predictive of actual usage.

Our study advances an investigation of datasets and publications with
novel ideas from network analysis. Our work puts together concepts
from other popular network flow estimation algorithms such as \textsc{PageRank}
\citep{brin1998anatomy} and \textsc{NetworkFlow} \citep{walker2007ranking}.
While scientists might potentially take a great amount of time to
change their citation behavior, we could use techniques such as the
one put forth in this article to accelerate the credit assignment
for datasets. Ultimately, the need for tracking datasets will only
become more pressing and therefore we must adapt or miss the opportunity
to make datasets first-class citizens of science.

\section*{Acknowledgements}

The authors would like to thank the Dr. Kim Pruitt from National Center
for Biotechnology Information, NLM, NIH, DHHS. Tong Zeng was funded
by the China Scholarship Council \#201706190067. Sarah Bratt was partially
funded by National Science Foundation award \#1561348. Daniel E. Acuna
was partially funded by the National Science Foundation awards \#1800956.

\bibliographystyle{apalike}
\bibliography{datarank_wu_bratt_zeng_acuna}

\end{document}